\documentclass[runningheads]{llncs}
\usepackage{comment}
\usepackage{graphicx}
\usepackage{amsmath}
\usepackage{multirow}
\usepackage{makecell}
\usepackage{amssymb}
\usepackage{booktabs}
\usepackage{utfsym}
\usepackage{cite}
\usepackage{xcolor}

\begin{document}
\titlerunning{Deep multimodal fusion of UWF-CFP and OCTA images}

\title{Improved Automatic Diabetic Retinopathy Severity Classification Using Deep Multimodal Fusion of UWF-CFP and OCTA Images}
%

\authorrunning{M. El Habib Daho et al.}

%

\author{Mostafa El Habib Daho \inst{1,2,*} 
\and
Yihao Li\inst{1,2,*} 
\and
Rachid Zeghlache\inst{1,2} 
\and
Yapo Cedric Atse \inst{1,2}
\and
Hugo Le Boité\inst{3,4}
\and
Sophie Bonnin\inst{5} 
\and
Deborah Cosette\inst{6} 
\and
Pierre Deman\inst{7,8} 
\and
Laurent Borderie\inst{8} 
\and
Capucine Lepicard\inst{9} 
\and
Ramin Tadayoni\inst{3,5} 
\and
Béatrice Cochener\inst{1,2,10} 
\and
Pierre-Henri Conze\inst{11,2} 
\and
Mathieu Lamard\inst{1,2} 
\and
Gwenolé Quellec\inst{2} 
}

\authorrunning{M. El Habib Daho et al.}
%
\institute{
Univ Bretagne Occidentale, Brest, France \and
LaTIM UMR 1101, Inserm, Brest, France \and
Ophthalmology department, Lariboisiere Hospital, APHP, Paris, France \and
Paris Cité University, Paris, France \and
Ophthalmology Department, Rothschild Foundation Hospital, Paris, France \and
Carl Zeiss Meditec Inc, Dublin, CA, United States
\and
ADCIS, Saint-Contest, F-14280 France
\and
Evolucare Technologies, Le Pecq, F-78230 France 
\and
AP-HP, Paris, France
\and
Ophthalmology Department, CHRU Brest, Brest, France \and
IMT Atlantique, Brest, France 
}

\maketitle              

\begin{abstract}
Diabetic Retinopathy (DR), a prevalent and severe complication of diabetes, affects millions of individuals globally, underscoring the need for accurate and timely diagnosis. Recent advancements in imaging technologies, such as Ultra-WideField Color Fundus Photography (UWF-CFP) imaging and Optical Coherence Tomography Angiography (OCTA), provide opportunities for the early detection of DR but also pose significant challenges given the disparate nature of the data they produce. This study introduces a novel multimodal approach that leverages these imaging modalities to notably enhance DR classification. Our approach integrates 2D UWF-CFP images and 3D high-resolution 6x6 mm$^3$ OCTA (both structure and flow) images using a fusion of ResNet50 and 3D-ResNet50 models, with Squeeze-and-Excitation (SE) blocks to amplify relevant features. Additionally, to increase the model's generalization capabilities, a multimodal extension of Manifold Mixup, applied to concatenated multimodal features, is implemented. Experimental results demonstrate a remarkable enhancement in DR classification performance with the proposed multimodal approach compared to methods relying on a single modality only. The methodology laid out in this work holds substantial promise for facilitating more accurate, early detection of DR, potentially improving clinical outcomes for patients.

\keywords{Diabetic Retinopathy Classification \and Multimodal Information Fusion \and Deep learning \and UWF-CFP \and OCTA}
\end{abstract}

\section{Introduction}
Diabetic Retinopathy (DR), a common ocular complication of diabetes, is a leading cause of blindness globally \cite{teo2021global}. The disease is characterized by progressive damage to the retina due to prolonged hyperglycemia and is estimated to affect approximately one-third of all people with diabetes. As such, timely and accurate diagnosis of DR is crucial for effective management and treatment. However, the subtle and complex nature of the disease's early stages presents a challenge for such a diagnosis.\\
Recent advances in imaging techniques have significantly enhanced the ability to detect and classify DR. Ultra-WideField Color Fundus Photography (UWF-CFP) imaging and Optical Coherence Tomography Angiography (OCTA) are two such techniques that have shown great promise. UWF-CFP imaging offers a panoramic view of the retina, allowing for a more comprehensive assessment \cite{Silva2015DR}, while OCTA provides depth-resolved images of retinal blood flow, revealing detailed microvascular changes indicative of DR \cite{Sun2021}.\\
Despite the individual merits of these imaging modalities, each offers a unique perspective on retinal pathology. Leveraging the information from both could potentially enhance the diagnosis and classification of DR \cite{Yang2021,Li2022UwfOcta}. However, the integration of these modalities poses a significant challenge due to the disparate nature of the data they produce, especially in terms of dimensionality (2D versus 3D) and field of view. \\
In recent years, deep learning (DL) has emerged as a powerful tool for medical image analysis, demonstrating great performance in a wide range of tasks\cite{li2021applications,quellec2021ExplAIn,Lahsaini2021Covid,Shamshad2023Transformers}. These models, particularly Convolutional Neural Networks (CNNs), have shown their ability to learn complex, hierarchical representations from raw image data, making them a natural choice for multimodal image fusion. \\
In the quest to enhance DL models, the field has benefitted significantly from incorporating innovative techniques like the Manifold Mixup \cite{verma2019manifold}. Through its unique method of generating virtual training examples via the convex combinations of hidden state representations, this technique has made a profound impact by significantly reducing a model's sensitivity to the data distribution and encouraging smoother decision boundaries.\\
Building upon these advanced techniques, several proposed methods in the state of the art have employed multimodal imaging \cite{Yihao2022Omia, Sleeman2023Multimodal}. These methods aim to utilize the complementary information available in different types of images. Recent works have effectively used mixing strategies to enhance multimodal DL models. For example, the M$^3$ixup approach \cite{Lin23} leverages a mixup strategy to enhance multimodal representation learning and increase robustness against missing modalities by mixing different modalities and aligning mixed views with original multimodal representations. The LeMDA (Learning Multimodal Data Augmentation) \cite{liu2023learning} method automatically learns to jointly augment multimodal data in feature space, enhancing the performance of multimodal deep learning architectures and achieving good results across various applications. MixGen \cite{hao2023mixgen} introduces a joint data augmentation for vision-language representation learning to boost data efficiency, generating new image-text pairs while preserving semantic relationships. This method has shown remarkable performance improvements across various vision-language tasks. Furthermore, TMMDA (Token Mixup Multimodal Data Augmentation) \cite{Zhao2023TMMDA} for Multimodal Sentiment Analysis (MSA) generates virtual modalities from the mixed token-level representation of raw modalities, enhancing representation learning on limited labeled datasets.\\
Despite the significant results obtained, these methods are proposed for vision-language and vision-audio fusion but are not suitable for 2D image/3D volume fusion. This study proposes a new multimodal DL approach for DR classification, integrating 2D UWF-CFP images and 3D OCTA images and incorporating a custom mixing strategy. Regarding the used modalities in this work, recent research has used UWF-CFP and OCTA imaging for the diagnosis of diseases such as Alzheimer \cite{Wisely2022Alzheimer}. However, to the best of our knowledge, our study is the first to develop a DL model for the classification of DR using both UWF-CFP and OCTA imaging modalities, which contributes significantly to the existing body of knowledge. \\

\section{Methods}

\subsection{Model architecture}
\begin{figure}[!t]
\centering
\includegraphics[width=\textwidth]{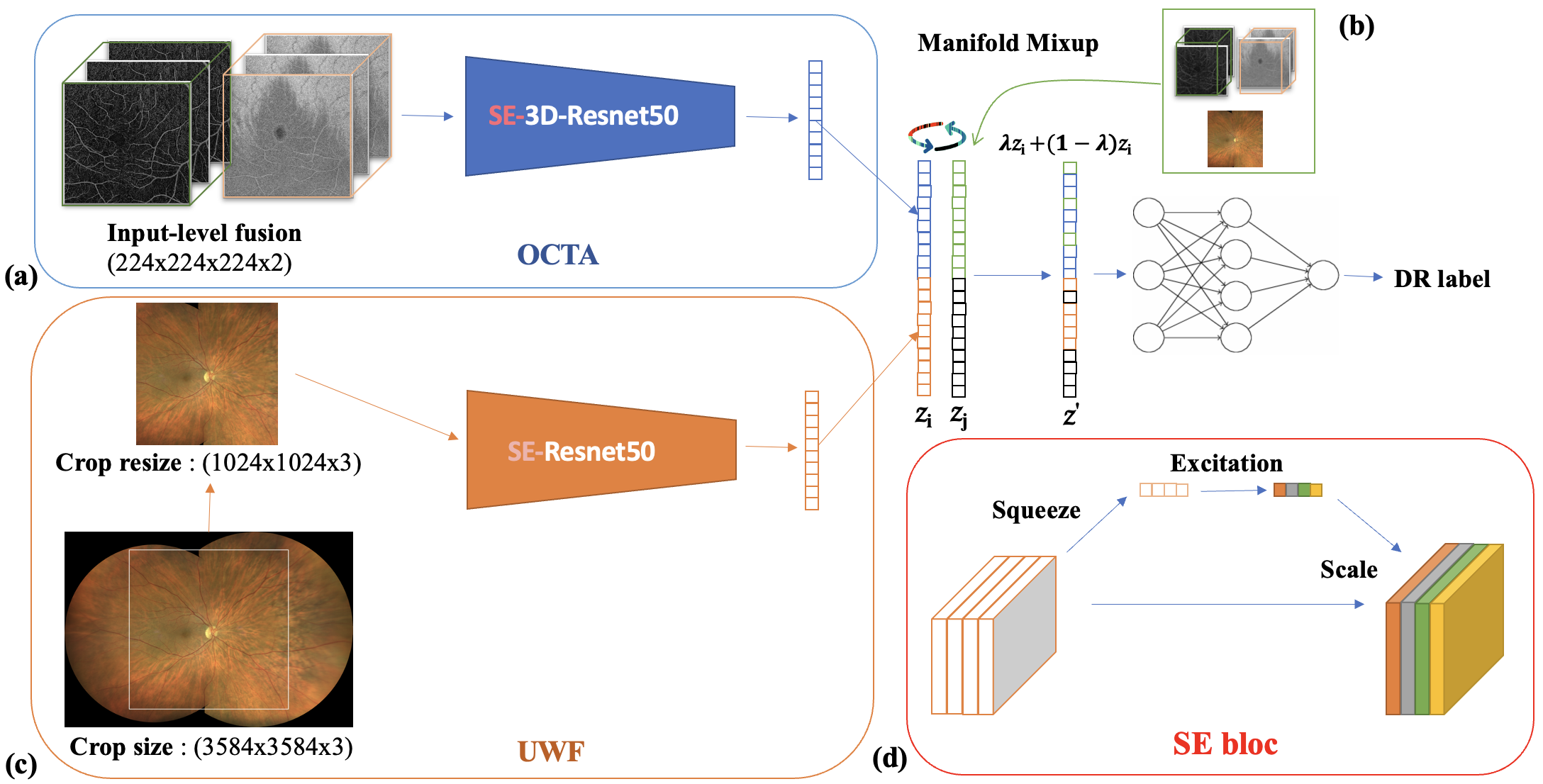}
\caption{Proposed pipeline.} \label{fig1}
\end{figure}

We utilize two separate CNN architectures, ResNet50 and 3D-ResNet50, designed to process 2D UWF-CFP and 3D OCTA images, to extract features from each imaging modality.
ResNet50 was chosen as a backbone for feature extraction due to its remarkable performance in computer vision tasks. Its structure provides a balance between depth and complexity, allowing the network to learn complex patterns without suffering from overfitting. To further improve such models' performance, Squeeze-and-Excitation (SE) blocks have gained attention in the DL community \cite{hu2018squeeze}. As shown in Fig.\ref{fig1}(d), the SE blocks dynamically recalibrate channel-wise feature responses by explicitly modeling the interdependencies between channels, thus helping the model focus on more informative features. They have been demonstrated to significantly improve the representational power of deep networks without a significant additional computational cost. \\
The 3D-ResNet50, a 3D extension of the ResNet50 architecture, integrated with SE blocks, is applied to process OCTA images (Fig.\ref{fig1}(a)). This model expands traditional 2D convolution operations into the 3D space, making it particularly appropriate for volumetric image data. This enables the network to decipher spatial hierarchies inherent in volumetric data, thus facilitating a comprehensive feature extraction from OCTA volumes. SE blocks in the 3D-ResNet50 model perform a similar role as in the 2D ResNet50 model, thus enhancing the performance of the 3D backbone. For the rest of the paper, we will refer to these models as SE-ResNet50 and SE-3D-ResNet50. 

\subsection{Fusion strategy}
The fusion of multiple modalities has been an area of active research due to the enhanced performances it offers \cite{akhavan2018combination,qian2020combined,zong2020deep}. Such fusion can be executed at input, feature, and decision levels, each offering distinct advantages and disadvantages.\\
In this work, we employ an input-level fusion for merging the structure and flow information embedded in OCTA images. Numerous studies affirm that merging these distinct types of information can significantly enhance the accuracy of DR diagnosis \cite{zang2022diabetic,Yihao2022Omia}. Input-level fusion involves integrating multiple modalities into a single data tensor subsequently processed by a DL model Fig.~\ref{fig1}(a). This method is effective without the need for registration, as the structure and flow data align with each other by design. \\

On the other hand, the fusion of UWF-CFP and OCTA images is performed through a different approach, primarily due to the absence of inherent alignment between these imaging modalities. Here, a feature-level fusion strategy is adopted, which allows us to use different backbones for each modality (SE-ResNet50 and 3D-SE-ResNet50), thus effectively addressing the alignment challenge. We have chosen feature-level fusion over decision-level fusion to capitalize on the rich interplay between the modalities at the feature level. This strategy facilitates the extraction of features and the fusion of high-dimensional feature-level information, making it especially suited for unregistered or dimensionally diverse data \cite{wu2022gamma,al2022cardiovascular,xiong2022multimodal,el2020multimodal}.

\subsection{Manifold Mixup}

To enhance the model's robustness and generalization capabilities, we implemented a multimodal extension of Manifold Mixup into our training process. The original Manifold Mixup method \cite{verma2019manifold} is a recently introduced regularization technique. It generates virtual training examples by forming convex combinations of the hidden state representations of two randomly chosen training examples and their associated labels.\\
Extending the concept of Input Mixup \cite{Zhang2017Mixup} to the hidden layers, Manifold Mixup serves as a robust regularization method that provokes neural networks to predict interpolated hidden representations with lesser confidence. It leverages semantic interpolations as an auxiliary training signal, leading to the cultivation of neural networks with smoother decision boundaries across multiple representation levels. Consequently, neural networks trained with Manifold Mixup can learn class representations with reduced directions of variance, thus yielding a model that exhibits enhanced performance on unseen data\cite{verma2019manifold}.
The operational process of the Manifold Mixup approach is as follows:

\begin{enumerate}
    \item The original Manifold Mixup performs the mixing of the hidden representation randomly on a set of predefined eligible layers. Instead, in our proposed implementation, we have purposefully selected the layer containing the concatenated feature maps from UWF-CFP and OCTA images to process the Manifold Mixup. This strategic choice is not only the simplest way to introduce Manifold Mixup but also ensures we are capitalizing on a layer that encapsulates a high-dimensional, multimodal feature space. Creating numerous virtual training samples from the fusion layer significantly improves the model's ability to generalize to new data. 

    \item Feed two images into the neural network until the selected layer is reached.
    \item Extract the feature representations ($z_i$ for multimodal data $x_i$ and $z_j$ for multimodal data $x_j$).
    \item Mix the extracted feature representations according to Eq.\ref{eq1} in order to derive the new representation (new features $z'$ associated with new label $y'$). 

    \begin{equation}
    (z', y') = (\lambda z_i + (1-\lambda) z_j, \lambda y_i + (1-\lambda) y_j)
    \label{eq1}
    \end{equation}
    
    where $z_i$ and $z_j$ are the features of two random training examples, and $y_i$ and $y_j$ are their corresponding labels. $\lambda \in [0,1] $ is a Mixup coefficient sampled from a Beta distribution $Beta(\alpha,\alpha)$, where $\alpha$ is a hyperparameter that determines the shape of the Beta distribution.
    
    \item Carry out the forward pass in the network for the remaining layers with the mixed data.
    \item Use the output of the mixed data to compute the loss and gradients. Given $\mathcal{L}$ the original loss function, the new loss $\mathcal{L}'$ is computed as:

    \begin{equation}
    \mathcal{L}' = \lambda \mathcal{L}(y_i, y') + (1 - \lambda) \mathcal{L}(y_j, y')
    \end{equation}

\end{enumerate}

Through this process, Manifold Mixup enhances our fusion strategy by operating on the joint feature representation (Fig.\ref{fig1}(b)), thereby ensuring that the model can generalize from the learned features of UWF-CFP and OCTA images.

\section{Experiments and Results}
\subsection{Dataset}
\setcounter{footnote}{0} 
The data used in this study arise from the "Évaluation Intelligente de la Rétinopathie diabétique" (EviRed) project\footnote{\url{https://evired.org/}}, a comprehensive initiative that collected data between 2020 and 2022 from 14 hospitals and recruitment centers across France. This database included UWF-CFP images and OCTA images from patients at various stages of DR. The dataset comprised images of 875 eyes belonging to 444 patients and was carefully divided into one (fixed) test set, and multiple train and validation sets (through 5-fold cross-validation) to ensure a broad representation and unbiased learning. Each patient's eye was labeled by an ophthalmologist into one of the 6 DR classes: Normal, mild nonproliferative diabetic retinopathy (NPDR), moderate NPDR, severe NPDR, proliferative DR (PDR), or Pan-Retinal Photocoagulation (PRP).\\
The UWF-CFP images in the dataset, captured using the Clarus 500 (Carl Zeiss Meditec Inc., Dublin, CA, USA), varied in size, ranging from 3900$\times$3900 to 7900$\times$4900 pixels. This size variation arises from the image stitching process for montage creation, not from changes in the device's resolution. Considering the clinicians' focus on the seven Early Treatment Diabetic Retinopathy Study (ETDRS) fields \cite{ETDRS1991}, we carried out center cropping on each image to 3584$\times$3584. This process ensured that all seven fields were included in the image. Subsequently, we resized these cropped images to 1024$\times$1024, a size that guarantees no loss of details.\\
The high-resolution 6x6 mm$^3$ OCTA images, offering 500$\times$224$\times$500 voxels and centered on the macula, were captured using the Zeiss PLEX Elite 9000. Each OCTA volume includes 2-D en-face localizer, structural, and flow 3D volumes. Due to the restrictions posed by the graphics processing unit (32Gb GPU) hardware, our 3D-SE-ResNet50 could only accommodate inputs up to $224\times224\times224\times2$ input tensors. This limitation guided our data pre-processing. In the training step of our deep learning network, we employed random crop processing. During the prediction process, we extracted multiple volumes from the OCTA image using N=10 times random crop, which were simultaneously processed with the full UWF-CFP image to make predictions. The final prediction for an examination was determined based on the severest prediction among these N predictions (test-time augmentation).

\subsection{Implementation details}
Our models were implemented using the PyTorch\footnote{\url{https://pytorch.org/}} deep learning library, and all experiments were conducted using an NVIDIA Tesla V100s GPU. For UWF-CFP images, we used the SE-ResNet50 architecture with weights pre-trained on ImageNet, while for OCTA images, we trained from scratch our implementation of the 3D-SE-ResNet50 backbone with input-level fusion for structure and flow volumes. The key to our model enhancement process included incorporating SE blocks in both ResNet models and using Manifold Mixup on multimodal features for model regularization. In our implementation, we set the reduction ratio, a crucial SE hyperparameter, to 16, following the practice from the original SE network paper \cite{hu2018squeeze}. For Mixup, we carried out a grid search focusing on the $\alpha$ parameter, which is essential for deriving the adequate Beta distribution $Beta(\alpha, \alpha)$ for sampling the right $\lambda$ interpolation parameter during Manifold Mixup training. This comprehensive exploration determined 0.2 as the optimal value for $\alpha$, which yielded the best model performance. The two models were trained jointly on the UWF-CFP and OCTA datasets, using a cross-entropy loss function and an AdamW optimizer. During training, we used a learning rate of 0.001 with the OneCycle scheduler, a decay factor of 0.0001, and a batch size of 4 over 200 epochs.

\subsection{Results and discussion}
To compare the performance of our proposed method with the individual modalities, we trained standalone models using either UWF-CFP or OCTA images with the same training settings as described above. This provided a baseline performance for each modality, against which the performance of the multimodal approach was compared. In addition, an ablation study was conducted to further understand each component's impact and contribution to our pipeline. We compared the performance of our model without the Manifold Mixup and the SE blocks. \\
The performance of the proposed models was evaluated in terms of the Area Under the Receiver Operating Characteristic (ROC) Curve (AUC). This metric was chosen due to its ability to provide an aggregate measure of performance across the four DR severity cutoffs ($\geq$ mild NPDR, $\geq$ moderate NPDR, $\geq$ severe NPDR, $\geq$ PDR). \\
Tab.\ref{table:1} presents the performance of the different models: the ResNet50 model trained on UWF-CFP images, the 3D-ResNet50 model trained on OCTA images, the proposed multimodal pipeline, the multimodal models without SE, the pipeline without Manifold Mixup (MM in the table), and the pipeline without SE and Manifold Mixup.

\begin{table}[h!]
\centering
\begin{tabular}{|c|c|c|c|c|c|c|}
\hline
\textbf{Data} & SE & MM &  $\geq$ mild NPDR & $\geq$ moderate NPDR & $\geq$ severe NPDR & $\geq$ PDR \\
\hline
UWF-CFP &\usym{2613}&\usym{2613}& 0.7983 &	0.7925 & 0.7906 & \textbf{0.9159} \\
OCTA &\usym{2613}&\usym{2613}& 0.8316 & 0.7627 & 0.7338 & 0.7576 \\
Multimodal & \checkmark &\checkmark  & \textbf{0.8566} & \textbf{0.8037} & \textbf{0.7922} & 0.8820 \\
\hline
\hline
Multimodal &\usym{2613} &\checkmark  & 0.8241 & 0.7969 & 0.7682 & 0.8522 \\
Multimodal &\checkmark  &\usym{2613} & 0.8431 & 0.7782 & 0.7566 & 0.8420 \\
Multimodal &\usym{2613} &\usym{2613} & 0.8140  & 0.7775 & 0.7525 & 0.8164 \\

\hline
\end{tabular}
\caption{Performance of Models in DR Classification}
\label{table:1}
\end{table}

Our approach that combines both UWF-CFP and OCTA images using a multimodal pipeline notably outperformed models based on individual modalities. Specifically, when evaluating DR severity cutoffs, the multimodal model achieved an AUC score of 0.8566 for $\geq$ mild NPDR, notably higher than 0.7983 for UWF-CFP alone and 0.8316 for OCTA alone. This trend continued with $\geq$ moderate NPDR and $\geq$ severe NPDR, where our multimodal model attained AUC scores of 0.8037 and 0.7922, respectively, compared to 0.7925 and 0.7906 for UWF-CFP and 0.7627 and 0.7338 for OCTA. These outcomes underscore the importance of capitalizing on diverse image modalities to provide a more comprehensive, holistic analysis, thereby enhancing the robustness and accuracy of DR classification. Our study suggests that each imaging modality captures distinct aspects of DR, and the concurrent utilization of both modalities in our models appears to improve the diagnosis, which is aligned with clinical studies \cite{Yang2021,Li2022UwfOcta}. \\
The greater success of UWF-CFP in identifying the cutoff $\geq$ PDR can be attributed to its wide-field view of the retina, which allows for the detection of peripheral lesions and signs of PRP laser impacts. Conversely, OCTA images proved to be particularly useful for $\geq$ mild NPDR detection due to their central focus on the macula and the high-resolution imaging of the microvasculature.\\
Regarding the added components in our pipeline, the Manifold Mixup and the SE blocks were proven to enhance the model's performance. For example, omitting the SE blocks caused a decrease in AUC scores across all DR severities. This indicates the critical role of SE blocks in bolstering feature representations and overall model robustness. Similarly, when the Manifold Mixup was excluded, there was a noticeable drop in performance, corroborating the effectiveness of such a regularization technique in improving model generalization.

\section{Conclusion}

Our findings demonstrate the efficacy of the proposed multimodal model in improving DR classification. This model, which integrates UWF-CFP and OCTA images using a feature-level fusion strategy and employing both our proposed adaption of the Manifold Mixup technique and SE blocks, delivers a compelling performance. The ablation study further attests to the significance of each component within our pipeline. These findings reiterate the necessity and potency of multimodal approaches coupled with advanced regularization techniques, such as Manifold Mixup and SE blocks, for medical image classification tasks. \\
To the best of our knowledge, our study is the first to propose a pipeline for the classification of DR using both UWF-CFP and OCTA images. However, we believe several improvements and extensions could further enhance the classification performance. The application of cross-modal attention mechanisms may provide a more effective way of fusing features from different modalities by focusing on the most relevant information from each. Similarly, implementing Manifold Mixup at different levels of the model, rather than solely at the concatenation layer, could provide further regularization and performance improvements. Moreover, introducing novel components, such as Transformer blocks, might prove beneficial in capturing complex relationships within and across modalities. 

\section*{Acknowledgements}
The work takes place in the framework of Evired, an ANR RHU project. This work benefits from State aid managed by the French National Research Agency under ``Investissement d'Avenir'' program bearing the reference ANR-18-RHUS-0008.

%
%
%

\bibliographystyle{splncs04}
\bibliography{Bibliography_omia23}

\end{document}